\documentclass[
    ,final            
  ]
  {aipproc}

\layoutstyle{6x9}

\begin{document}

\title{The \mbox{\boldmath $\tilde{U}(12)$}-Classification Scheme, Static $U(4)$-Spin 
Symmetry for Light-Quarks and ``Exotic'' Hadrons}

\classification{}
\keywords      {}

\author{Shin Ishida\footnote{Representing the collaboration group with M.Ishida, 
T Maeda, M. Oda and K.Yamada}}{
  address={Research Institute of Science and Technology, 
College of Science and Technology, \\
Nihon University, Tokyo 101-8308, Japan}}
\begin{abstract}
  Several years ago we have proposed a manifestly covariant 
  $\tilde{U}(12)_{SF}$-classification scheme of hadrons, which 
  maintains the successful $SU(6)_{SF}\otimes O(3)_{L}$ framework 
  in Non-Relativistic Quark Model and is also reconcilable with 
  the mechanism of Spontaneously-Broken Chiral Symmetry in 
  Relativistic Field Theoretical Models. The essential point here 
  is to notice the overlooked freedom of the $SU(2)_{\rho}$-spin 
  for (confined) light quarks, which 
  leads to existence of new type of ``exotic'' hadrons, called 
  ``chiralons'' and to a selection rule, $\rho_{3}$-line rule, on 
  the spectator quark line.
  
  A series of puzzling new hadrons recently observed such as 
  $X(3872)$-meson family and $\Theta (1540)$ and so on are 
  possibly classified mostly as chiralons.
\end{abstract}
\maketitle
\begin{figure}[t]
 \includegraphics[width=.7\textwidth, height=.2\textheight]{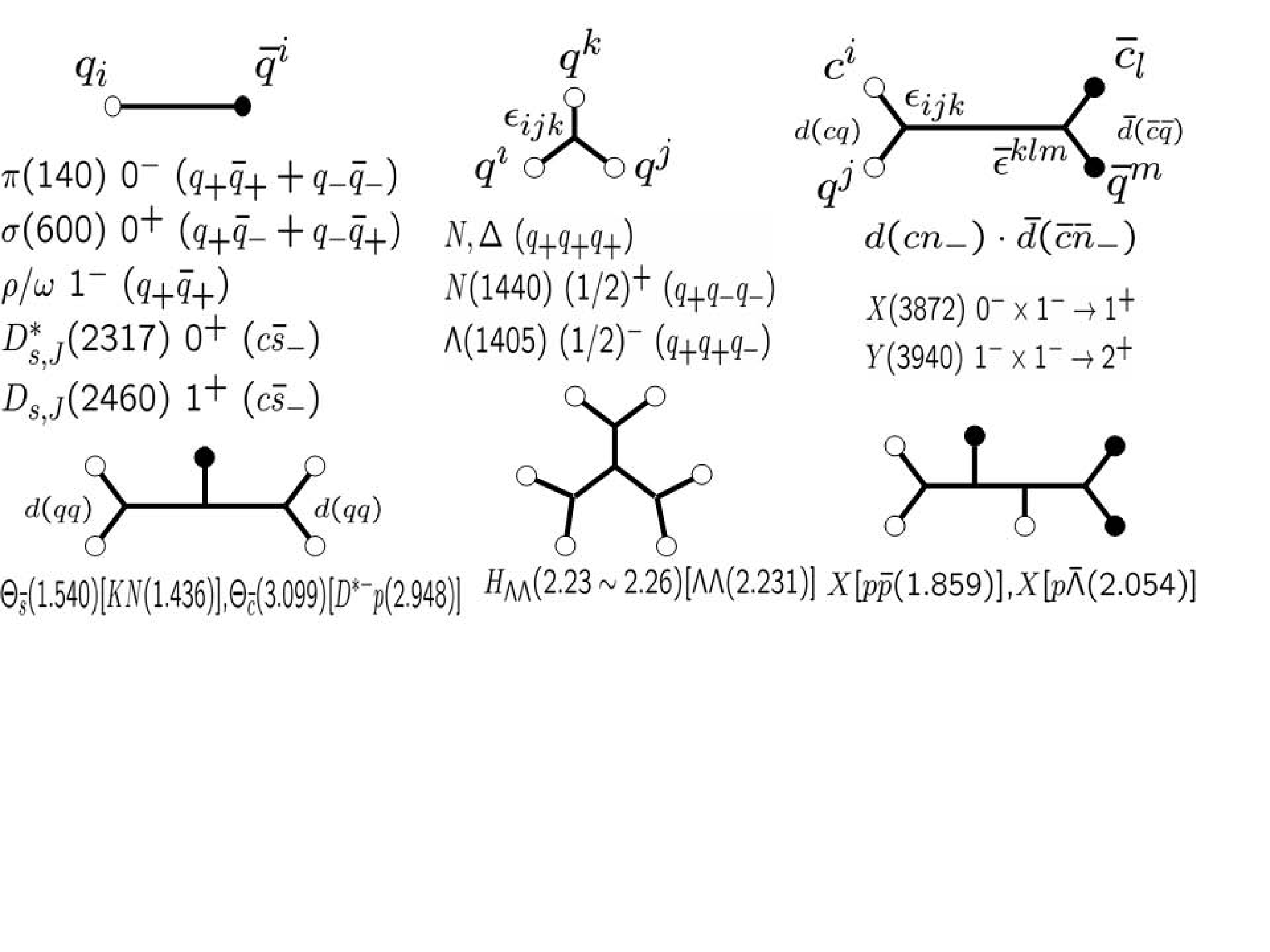}
\caption{Genaral hadrons in $\tilde{U}(12)$-classification scheme}
\label{fig:1}
\end{figure}

\vspace{-.5cm}
\begin{center}
{{\bf PRESENT STATUS ON HADRON SPECTROSCOPY}}
\end{center}
\ \ \ {\it (Conventional Two View-Points on Hadron Classification) } \ \ 
The non-relativistic view is based upon NRQM and has the successful 
$SU(6)_{SF}\otimes O(3)_{L}$ framework; while the relativistic view 
resorts to RFTM and realizes the mechanism of SBCS, 
an indispensable notion in the low-energy hadron physics. On the other hand, 
both have their own serious difficulties; the former is lacking Lorentz 
covariance and out of notion chiral symmetry; while the latter is unable 
to treat internal excitation of hadrons.\\
\ \ \ {\it (Proposal of $\tilde{U}(12)_{SF}$-Classification Scheme)}\ \ 
This[1] is a manifestly-covariant extension of NRQM in conformity with 
the chirality $\gamma_{5}$-transformation on the light quarks, and 
accordingly is reconciled with the mechanism of SBCS. The scheme has a 
unitary symmetry, $U(12)_{SF}\otimes O(3)_{L}$, 
in the rest-frame of hadrons, represented in the covariant 
$\tilde{U}(12)_{SF}\otimes O(3,1)_{L}$ space; that is, 
the hadron Wave Function, which is tensors 
in the above space, has the static symmetry 
$U(12)_{SF}\supset SU(3)_{F}\otimes U(4)_{S}$, 
embedded in $\tilde{U}(12)_{SF} \supset SU(3)_{F}\otimes \tilde{U}(4)_{DS}$, 
where  $U(4)_{S}\supset$ \fbox{$SU(2)_{\rho}$}$\otimes SU(2)_{\sigma}$ 
and the $U(4)_{S}$($\tilde{U}(4)_{DS}$) being 
a unitary group(pseudo-unitary Lorentz group) on the 4 components of 
Dirac spinor(the $\gamma$-matrices$\equiv \sigma\otimes\rho$). 
The basic vectors of the ``new'' freedom of $SU(2)_{\rho}$ are 
given by two eigenstates of $\rho_{3}$ with respective eigen value $r=\pm$; 
``Pauli-spinor'' $\Phi_{+}(X)$ with $j^{P}={\frac{1}{2}}^{+}$, 
and ``Chiral-spinor'' $\Phi_{-}(X)$ with $j^{P}={\frac{1}{2}}^{-}$. 
The chiral spinor $\Phi_{-}(X)$ with ``exotic '' quantum numbers 
leads to existence of ``exotic-hadrons''\footnote{We had defined the 
two kinds of hadron states: The chiral states are described 
by tensors with at least one chiral spinor-index, while the Pauli states 
are by those with only Pauli spinor-index. We call especially the hadrons 
represented purely by the Pauli/Chiral states as Paulons/Chiralons.}.\\
\ \ \ {\it (Observation of New Hadrons)}, \ \ seemingly to be out of the 
conventional classification scheme, has recently been reported successively. 
They have features : 
(F1) Their mass is mostly close to the threshold of their observed or 
intermediate decay channels, and (F2) their decay width is unexpectedly narrow.
The purpose of this work is to point out that these ``new hadrons'' 
be promising candidates 
of chiralons.
\begin{center}
{\bf{OVERVIEW OF HADRONS IN $\tilde{U}(12)$-SCHEME}}\\[.1cm]
\end{center}
\ \ \ {\it (Schematic Structures of General Hadrons)} in 
$\tilde{U}(12)$-classification scheme are shown in 
Fig.{\ref{fig:1}} systematically for conventional 
$(q\bar{q})$-mesons and $(qqq)$-baryons and also for 
multi-quark hadrons resorting on Joined-Spring Quark Model. 
It had been proposed so as to give color-singlet and triality-zero hadrons 
a long time ago.
In this Fig. 
we have, in addition to the internal 
space-time structure, 
given the structure on flavor, 
color and the new $SU(2)_{\rho}$-spin 
freedom of constituent quarks(, where necessary), 
and also made some tentative assignment of relevant new hadrons, 
taking into account theoretical 
expectations from the static $U(4)$-spin symmetry.\\
\ \ \ {\it (Existence of Chiral States and Tentative Assignments)}
The $U(4)_{S}$-spin WF of hadrons contains a relevant tensor product 
of basic vectors in $SU(2)_{\rho}$-space. 
The basic-vectors are two kinds of Dirac spinors, which are the Fourier conjugates of 
$\Phi_{\pm}(X)$, with definite four-volocity $v_{\mu}=P_{\mu}/M$ of relevant 
hadrons(, see later sections). 
In JSQM(where the symmetry due to quark statistics 
is somewhat restricted by connected springs) the relevant hadron WF 
is represented by those of 
quarks and of di-quarks with the quantum numbers $j^{P}$ as 
\vspace{-.2cm}
\begin{eqnarray}
q_{\pm}:{{1}/{2}}^{\pm}; \ \ \
d(q_{+}q_{+});0^{+}, 1^{+}; \ \ 
d^{\chi}(q_{+}q_{-});0^{-}, 1^{-}; \ \ {\rm{and}} \ \ 
d^{\chi\chi}(q_{-}q_{-});0^{+}, 1^{+}.
\end{eqnarray}
The existence of chiral spinor $q_{-}$ with $j^{P}={{1}/{2}}^{-}$ 
is the origin of exotics and leads to chiral states which show anyhow 
exotic properties.\\
\underline{WF of $(q\bar{q})$-mesons and $(qqq)$-baryons} 
The $SU(2)_{\rho}$-spin WF of [$\pi(140)$]/[$\sigma(600)$] nonets 
comes from their being linear representation of S.B. chiral symmetry. 
Similarly WF of $D_{s J}^{*}(0^{+};2317)$/$D_{s J}(1^{+};2460)$ reflects 
that they are chiral partners of $D_{s}(0^{-};1968)$/$D_{s}^{*}(1^{-};2112)$. 
The WF of  [$\rho$]/[$\omega$] being Paulons is due to their 
electro-magnetic properties. The WF for Roper resonance $N(1440)$ 
and $SU(3)$-singlet $\Lambda (1405)$ is supposed because of their 
comparatively lighter mass. \\
\underline{Multi-quark ``Exotic hadrons''} 
The $X(3872)$ family($X(3872)$,$Y(3940)$,$\cdots$) 
are considered to be good candidate of chiralons in the system of 
[$(cn_{-})\cdot(\overline{cn_{-}})$] tetra-quarks. The $X(3872)$ had 
the mass close-to-threshhold of its decay chanel, $J/\psi + \rho,\omega$, 
and unexpectedly small width $\Gamma < 2.3$MeV. 
In Fig. 
we have picked up also the other observed 
new hadrons with properties of \underline{close-to-threshold mass} 
and of the \underline{smaller-than-expected width}. 
\begin{center}
{\bf{OVERLOOKED FREEDOM OF $SU(2)_{\rho}$-SPIN}}\\[.1cm]
\end{center}
\ \ \ {\it (Summary of Covariant Description of Composite Hadrons)}
It should be first noted that we are not treating a dynamical 
bound-state problem but presenting a kinematical framework, 
where the c-number Dirac-spinor as a mathematical tool, called urciton,
is simulating the role of physical quarks.\\
\ \ We set up the hadron WF which transforms 
as a relevant tensor product of the c-number quark field 
and its Pauli conjugate, $\psi_{A}(x)$ ans 
$\bar{\psi}^{B}(x)$ ($A=({\alpha},a)$etc. $\alpha$($a$) 
denotes Dirac spinor(flavor) index), and 
start from the Yukawa-type 
Klein-Gordon equation to be satisfied by WF
\vspace{-.2cm}
\begin{eqnarray}
[ (\frac{\partial}{\partial X_{\mu}})^{2} 
- {\cal M}^{2} (r_{\mu},\partial/\partial r_{\mu} ) ]
\Phi_{A_{1}\cdots}^{B_{1}\cdots}(X, r, \cdots) =0, 
\label{eq:3}
\end{eqnarray}
where the ${\cal M}^2$ is 
depending on relative 
coordinates of constituents 
($X_{\mu}$/$r_{\mu}$ being C.M./relative coordinates), 
and assumed to be 
diagonal on 
$A_{1}$ and $B_{1}$etc. 
In this talk we concern only ground states 
neglecting the dependence on $r_{\mu}$ of WF. 
Then WF of hadrons $\Phi(X)$ is given as solutions of 
the ``local'' K.G.Eq.(\ref{eq:3}), 
and the positive(negative)-frequency Fourier conjugates of WF, 
become, 
through the second-quantization, to represent systematically 
the annihilation(creation) operator of all hadrons(anti-hadrons) 
composed in the relevant multi-quark system. 
The Fourier conjugates of Pauli-conjugate WF 
become similarly the creation(annihilation) operators. 
This leads to the crossing relation 
or substitution law of hadrons, an important attribute of hadrons. \\
\ \ The $\tilde{U}(4)_{DS}$ WF of hadrons with
 $v_{\mu}=P_{\mu}/M$ ($P_{\mu}$($M$) ;hadron momentum(mass)) 
is a tensor product of basic vectors, urciton-spinors, 
in $SU(2)_{\rho}$-space. They are defined through the 
two (Pauli and chiral) types 
of Dirac spinor $\Phi_{\alpha}(X)=\{ \Phi_{+\alpha}, \Phi_{-\alpha}(X) \}$
\footnote{
Whether the chiral spinor is required in the actual spectrum is 
judged only through comparison 
with experiments. 
However, for the confined light quarks inside of the lower mass hadrons 
it is indispensable to describe the relevant chiral symmetry. 
In describing hadrons with them 
both ``mass conjugate'' members of 
$W(P)$ and of $\bar{W}(P)$ 
are required, since the hole theory, where $U_{-}(P)=V_{+}(P)$ and 
$V_{-}(P)=U_{+}(P)$, is not applicable to the confined quarks 
independently of the remainders(, Note that particle and 
anti-particle have mutually conjugate-quantum numbers, for example, 
$q$($\bar{q}$) belongs to ${\underline 3}_{c}$(${\underline 3}_{c}^{*}$) on the color). 
On the other hand it is not the case for heavy-quarks. It may be 
evident from Eq.(4) that the $(\pm)$ sign of $\Phi_{\pm}$ becomes equal to that of 
$\rho_{3}$-spin in hadron frame.} 
, defined by the solutions 
of K.G. equation as 
\vspace{-.2cm}
\begin{eqnarray}
(\frac{\partial^2}{\partial {X_{\mu}^2}}-M^2)\Phi_{\alpha}(X)&=&[(\gamma_{\mu}\partial_{\mu}+M)( \gamma_{\mu}\partial_{\mu}-M)\Phi(X)]_{\alpha} =0,\\
( \gamma_{\mu}\partial_{\mu}+|M|)\Phi_{+} (X)&=&0; \  \  \ 
( \gamma_{\mu}\partial_{\mu}-|M|)\Phi_{-} (X)=0 \nonumber
\label{eq:5}
\end{eqnarray}
They are related mutually as 
$\Phi_{\pm,\alpha}(X)=-\gamma_{5} \Phi_{\mp,\alpha}(X)$, 
and the complete set of the basic vectors 
$W_{\alpha}(P)= \{ U_{\pm \alpha}(P) \}$ 
and ${\bar W}^{\beta}(P)=\{  {\bar V}_{\pm}^{\beta}(P) \}$ 
are given by Fourier conjugates of $\Phi(X)=\{ \Phi_{\pm}(X) \}$ 
and, as those of their Pauli conjugate $\bar{\Phi} (X)$. 
\vspace{-.2cm}
\begin{eqnarray}
\Phi_{\pm,\alpha}(X)\equiv {\sum}_{P_{\mu}(P_{0}>0)} 
{U}_{\pm\alpha}(P_{\mu})e^{iPX}, \ \ \ 
\bar{\Phi}_{\pm}^{\beta}(X)\equiv 
{\sum}_{P_{\mu}(P_{0}>0)} \bar{V}_{\pm}^{\beta}(P_{\mu})e^{iPX}, \ \ \ 
\label{eq:a}
\end{eqnarray}
where $(\bar{V}_{\pm}{}^{t})^{\beta}(P_{\mu})=\mp C^{\beta\alpha} {U}_{\pm\alpha}(P_{\mu})$ 
($C=\gamma_{4}\gamma_{2}=-i\rho_{1}\sigma_{2}$ is charge- conjugation matrix).
\begin{center}
{\bf{MULTI-QUARK EXOTIC HADRONS AND $X(3872)$ FAMILY}}\\[.1cm]
\end{center}
\ \ \ {\it (Properties of Multi-Quark Hadrons)}
The mass of ground states is generally given as: 
$M=M^{(0)}+\delta M$; $M^{(0)}
= \sum_{i} m_{i}$(sum of constituent masses), 
where $M^{(0)}$ is mass in the ideal limit 
(being the $U(12)_{SF}$-symmetric on light-quarks), 
while $\delta M$ represents the effects of broken 
chiral-symmetry($\delta^{\chi} M$) and of perturbative QCD($\delta^{OGE} M$). 
Accordingly 
the total mass $M^{(0)}$ is degenerate through all 
types of hadron and/or hadron systems at their 
thresholds if they as a whole have the same quark-flavor 
configurations. This leads to F1 one feature of puzzling new hadrons. 
For an example of $X(3872)$, 
\vspace{-.2cm}
\begin{eqnarray}
M^{(0)}_{T}((cq)(\overline{cq}))= M^{(0)}_{\psi} (c\bar{c}) + M^{(0)}_{M} (q\bar{q})
=M^{(0)}_{D} (c\bar{c}) + M^{(0)}_{\bar{D}} (\bar{c}q)
\end{eqnarray}
Now we focus our attention 
on the fission process with small phase volume, as in the above example, 
among ground state hadrons.
In this process the transition matrix element $M$ is considered to 
come purely from the overlapping $I_{O.L.}$ of WF as 
$M=g^{(0)} \times I_{O.L.}$
($g^{(0)}$; a dimensional 
parameter), and $I_{O.L.}$ consists of a product of 
those on each quark line as 
$I_{O.L.}=\Pi_{i} \ I_{O.L.} [q_{i}]$. 
The $I_{O.L.} [ q ] $ is just the inner product between WF 
of initial and final quarks. Then, because of the static $U(4)$-spin symmetry 
for light-quarks, the transiton between urcitons with different r-value is 
forbidden at threshold, while suppressed strongly in the region close to 
threhold. This, a kind of selection rule to be called $\rho_{3}$-line rule, 
leads to F2, another feature of new hadrons. The suppression factor is, 
concerning the $SU(2)_{\rho}$-space, $\epsilon ({\mathbf P})=|{\mathbf P}|
/(M+E(|{\mathbf P}|))<1$(${\mathbf P}$, E, and $M$ 
being momentum, energy, and mass of the relevant final hadrons). 
In this connection the example of $X(3872)$-family are instructive: Firstly 
the mass of $X(3872)$ is almost equal to the threshold of ($D(1870)$, 
$\bar{D}^{*}(2010)$) and is within the threshold region of ($\rho(\omega)$, 
$J/\psi$). Both the decay channels are doubly forbidden by the 
$\rho_{3}$-line rule. Experimentally the decay into the latter channel has 
been obseved with a small width $\Gamma< 2.3$MeV, but not into the former. 
This fact is understood by the $\rho_{3}$-line rule, considering the relation 
between respective suppression factors, 
$\epsilon^{4}({\mathbf P})_{D\bar{D}^{*}}
<\epsilon^{4}({\mathbf P})_{J/\psi\cdot \rho(\omega)}\ll 1$. 
Secondly the mass of $Y(3.943)$ is well above the open-charm thresholds
(3.872MeV) of $D\bar{D}^{*}$ and near to the threshold(3.880) of 
($\omega$, $J/\psi$), of which both decay-channels are similarly 
double-forbidden by $\rho_{3}$-line rule. Experimentally its decay into the 
$\omega J/\psi$ has been observed with a rather narrow width 
$\Gamma\approx 90$MeV, while not into the channel $D\bar{D}^{*}$. 
This situation may be also understandable from the difference of 
the suppression factors, $\epsilon_{D\bar{D}^{*}}
<\epsilon_{J/\psi\cdot \omega}$.\\
\ \ \ {\it (Level Structures of Tetra-quark$[(cq)(\overline{cq})]$-System)}
The quantum numbers of groundstate multiplets are obtained from those of 
di-quarks and anti-diquarks, $d(cq)$ and $\bar{d}(\overline{cq})$
(see, Eq.(1)). The relevant $SU(2)_{\rho}$-space WF are classified into 
three-groups; $T_{\chi\chi}(d^{\chi}\bar{d}^{\chi})$, 
$T_{P\chi}(d^{\chi}\bar{d}^{P}/d^{P}\bar{d}^{\chi})$ and 
$T_{PP}(d^{P}\bar{d}^{P})$(where $d^{\chi}\equiv (cq_{-})$ and 
$d^{P}\equiv (cq_{+})$) with $j^{P}(T)=
j^{P}(d)\otimes j^{P}(\bar{d})$). In order to obtain the more realistic 
mass-spectra is necessary to estimate the symmetry breaking effects 
$\delta M = \delta^{\chi} M+ \delta^{OGE} M$, which are approximated by 
the respective sum of those of constituents, 
$\delta M_{T}=\delta M_{d}+\delta M_{\bar{d}}$. \\
\ \ Concerning $\delta^{\chi} M \equiv M^{\chi}-M^{P}$, 
applying the L$\sigma$M 
for the Yukawa interaction of light quarks inside of $d(cq)$-diquarks 
and of $D(c\bar{q})$ mesons, we get the relation, 
\vspace{-.2cm}
\begin{eqnarray}
\delta^{\chi} M_{d}(cq)&=&-\delta^{\chi} M_{D}(c\bar{q}) \ \ \ 
(\delta^{\chi} M_{D}(c\bar{n})\propto a, \ \delta^{\chi}M_{D}(c\bar{s}) \propto b)
\end{eqnarray}
where $a\equiv \langle n \bar{n} \rangle_{0}$ and 
$b\equiv \langle s\bar{s} \rangle_{0}$ are vacuum eqpectation values of 
respective scalar densities. Using the experimental values of 
$\delta^{\chi} M_{D}(c\bar{s})$ and  of the ratio $a/b$ we obtain[2] 
the values; 
$\delta^{\chi} M_{d}(cn)=-242$MeV, 
$\delta^{\chi} M_{d}(cs)=-348$MeV. This implies that the masses of 
$T[(cq)(\overline{cq})]$ -system are in order of $M_{T}(\chi\chi)<
M_{T}(P\chi)<M_{T}(PP)$ with the difference $\delta^{\chi} M_{d}(cq)$, 
Eq.(6).
It is notable that the negative 
relative-sign between both sides of Eq.(6) reflects the Charge-Conjugation 
property of thr c-number scalar-density. \\
\ \ Concerning $\delta^{OGE} M$, 
we set up the effective spin-current interaction 
between c-quarks and light-quarks as 
\vspace{-.2cm}
\begin{eqnarray}
\langle H^{J} \rangle \approx \langle \bar{c} \sigma_{\mu\nu} c \rangle 
\langle \bar{q}_{\pm} F_{U} \sigma_{\mu\nu} q_{\pm} \rangle
\propto \langle {\mathbf \sigma}^{(c)} \rangle
\langle {\mathbf \sigma}^{(q)} \rangle;
\end{eqnarray}
where the middle term is color-gauge invariant and becomes static $U(4)_{S}$
-symmetric  due to the role of unitarizer $F_{U}$. This leads to the 
relation of $\delta^{J} M \equiv M(J=1)- M(J=0)$, 
\begin{eqnarray}
\delta^{J} M_{d}(cq)=({1}/{2}) \delta^{J} M_{D}(c\bar{q}) =71
{\rm MeV  \ \ \ for \ \ } q=(n,s)
\end{eqnarray}
where the positive relative-sign of the first equality in Eq.(7) reflects the 
C.C. property of the c-number tensor-density and the factor 1/2 comes from 
the color freedom. The value of Eq.(8) is obtained from the 
experimental values in the $D(c\bar{q})$ meson system. 
 Here it may be notable that the mass defference between $X(3872)$ and $Y(3940)$ is 
close to the estimated one Eq.(8). The more detailed investigations on $X(3872)$ family 
are reported in this conference[3]. 
Finally, we give brief comments on the related works: 
The diquark and anti-diquark picture of $T$-quark system 
was first discussed in Ref.[4], but without considerations on 
chiral symmetry. Actually the relevant level structure of [4] is 
equivalent to that of our $T_{PP}$ system. The $D^{0}\bar{D}^{*0}$ 
molecular picture[5] also explains(or starts from) the feature 
of new hadrons(F1). However, its WF has very loose structure whereas in our case 
it is, supposed to be, tight due to color confining force.
\bibliographystyle{aipproc}   
\begin{center}
{\bf{REFERENCES}}\\
\end{center}
\small
1. S.~Ishida and M.~Ishida, {\it Phys. Lett.}\textbf{B539}, 249(2002); 
S.~Ishida, M.~Ishida, and T.~Maeda, {\it Prog. Theor. Phys.}\textbf{104}, 
785(2000).\\
2. S.~Ishida et al.,hep-ph/0408136; See, also the precedent work, 
W. A. Bardeen, E. J. Eichten and C. T. Hill, 
{\it Phys. Rev.}\textbf{D68}, 054024(2003).\\
3. M.~Ishida, in proc. of HADRON'05.\\
4. L.~ Maiani, F. Piccinini, A. D. Polosa and V. Riquer, {\it Phys. Rev.}\textbf{D71}, 014028(2005).\\
5. F. E. Close and P. R. Page, {\it Phys. Lett.}\textbf{B578}, 119(2004); 
N. A. Tornqvist, {\it Phys. Lett.}\textbf{B590}, 209(2004); 
C. Y. Voloshin, {\it Phys. Lett.}\textbf{B579}, 316(2004); 
E. Braaten and M. Kusunoki, {\it Phys. Rev.}\textbf{D69}, 074005(2004); 
E. S. Swanson, {\it Phys. Lett.}\textbf{B588}, 189(2004); \textbf{B598}, 197(2004)
\end{document}